\documentclass{PoS}
\usepackage{graphicx}
\usepackage{listings}

\title{mc4qcd: Online Analysis Tool for Lattice QCD}

\ShortTitle{mc4qcd}

\author{\speaker{Massimo Di Pierro}\\
        School of Computing - DePaul University - Chicago\\
        E-mail: \email{mdipierro@cs.depaul.edu}}

\author{Yaoqian Zhong\\
        School of Computing - DePaul University - Chicago\\
        E-mail: \email{ati\_zhong@hotmail.com}}

\author{Brian Schinazi\\
        School of Computing - DePaul University - Chicago\\
        E-mail: \email{schinazi@gmail.com}}

\abstract{mc4qcd is a web based collaboration tool for analysis of Lattice QCD data. Lattice QCD computations consists of a large scale Markov Chain Monte Carlo. Multiple measurements are performed at each MC step. Our system acquires the data by uploading log files, parses them for results of measurements, filters the data, mines for required information by aggregating results, represents the results as plots and histograms, and it further allows refining and interaction by fitting the results. The system computes moving averages and autocorrelations, builds bootstrap samples and bootstrap errors, and allows modeling the data using Bayesian correlated constrained linear and non-linear fits. It can be scripted to allow real time visualization of results form an ongoing computation. The system is modular and it can be adapted to automating the analysis workflow of different types of MC computations.}

\FullConference{13th International Workshop on Advanced Computing and Analysis Techniques in Physics Research, ACAT2010\\
		February 22-27, 2010\\
		Jaipur, India}

\begin{document}

\lstset{ %
basicstyle=\footnotesize,       
numbers=left,                   
numberstyle=\tiny,      
stepnumber=1,                   
numbersep=5pt,                  
showspaces=false,               
showstringspaces=false,         
showtabs=false,                 
frame=single,                   
tabsize=2,                      
breaklines=true,                
breakatwhitespace=false,        
}

\section{Introduction}

\begin{figure}
\begin{center}
\begin{tabular}{c}
\includegraphics[width=3.3in]{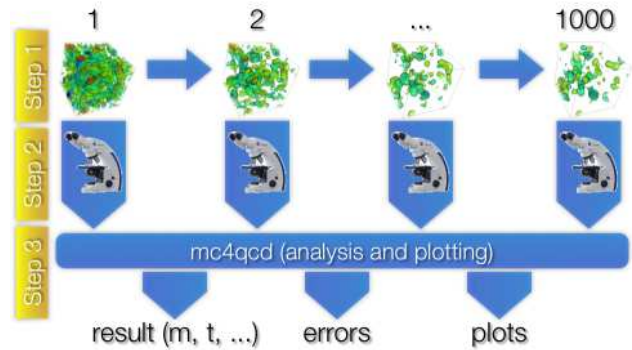}
\end{tabular}
\end{center}
\caption{The typical three steps of a lattice computation.
\label{fig1a}}
\end{figure}

Lattice QCD is a numerical approach to Quantum Chromodynamics.  Its primary goals are to describe the non-perturbative behaviour of quarks and to compute properties of hadronic matter. Typical Lattice QCD computations are comprised of three steps (~fig.\ref{fig1a}). Step 1 consists of a Markov Chain Monte Carlo that generates {\it gauge configurations}, i.e. datasets that represent possible evolutions of the gluonic field. Step 2 consists of performing measurements on each of these gauge configurations, typically measurements of correlation functions between different operators at different locations in space and time. Step 3 consists of averaging the results of step 2 and performing a statistical analysis of the result. The outputs of step 3 may include values such as the masses and lifetimes of mesons and baryons.

In this procedure, Step 1 is the most computationally expensive, but it is also the most automated. Step 2 is also computationally intensive, of magnitude dependent on the type of question that one is trying to answer. Typically, multiple measurements are performed on each gauge configuration such that one can answer multiple questions by aggregating them in different ways in Step 3. This final step is relatively inexpesive from a computational point of view, but it is where most physicists spend their time: analyzing data and generating plots. The mc4qcd software aims to automate this process.

\begin{figure}
\begin{center}
\begin{tabular}{c}
\includegraphics[width=4.5in]{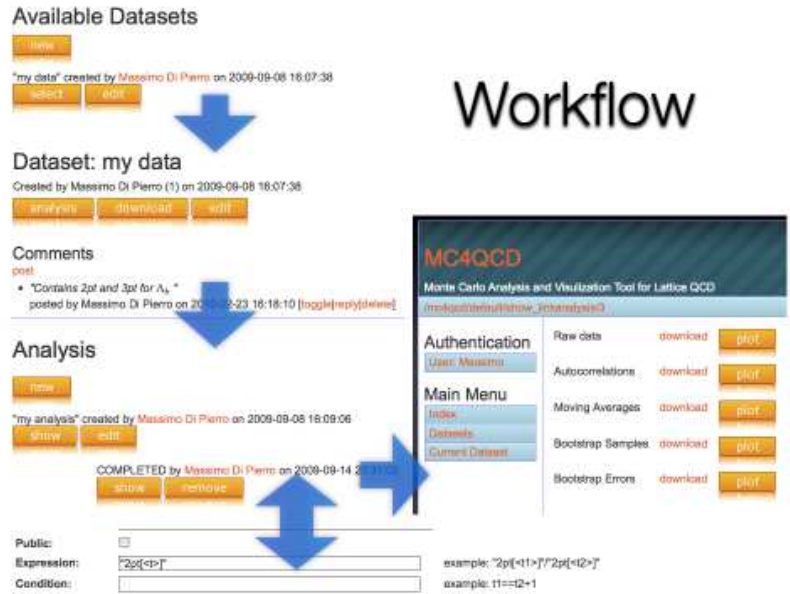}
\end{tabular}
\end{center}
\caption{A collection of screenshots of the mc4qcd program.
\label{fig1b}}
\end{figure}

The data communicated between Step 1 and Step 2 is highly stuctured, using one of a small set of file formats that scientists have agreed on. Conversely, there is no standrard format for encoding the data communicated between Step 2 and Step 3. Such standardization would be very difficult because the data is unstructured and is rarely communicated between different research groups. Additionally, the data is normally stored in flat text files with no metadata, save occasional comments printed in the output that are meant for human, not machine, consumption.

The mc4qcd application is specifically designed to ameliorate this situation.

The software performs the following basic operations:

\begin{itemize}\addtolength{\itemsep}{-0.5\baselineskip}

\item Import: Data can be uploaded either via a web interface or by means of web services.
\item Extraction and transformation: The application parses the output text files from Step 2 computations and uses regular expressions to identify important variables. The regular expressions provide a template for transforming the original files into a structured dataset (in the form a collection of tables).
\item Filtering: In addition to regular expressions, the user can apply mathematical constraints to filter data.
\item Mining: mc4qcd aggregates data by computing expressions of interest and performing standard statistical analysis.
\item Visualization: It produces thos plots that physicists normally look at to better read the information, including moving averages, autocorrelations, and bootstrap error.
\item Exploration: It allows users to explore their data by zooming plots and interactively applying addition filters.
\item Fitting: mc4qcd performs non-linear, constrained, correlated, and Bayesian fits, or combinations thereof.
\item Output: Data, results, and plots can be exported.
\item Administration: The software provides several administrative features, including web-based collaboration and commenting, traceability, ownership and permissions, public and private datasets, logging, storage of results, and annotations using LaTeX.
\end{itemize}

None of the above operations are new or exceedingly original. Physicists already perform them using other programs and plotting tools. The limitation of the current solutions is that they are not sufficiently general, and they require {\it ad hoc} programming.

The added value of mc4qcd lies in its generality, automation capabilities, ease of use, collaborative features, and its enforcement of a workflow (fig.~\ref{fig1b}).

\section{Architectire}

mc4qcd is written in Python. It consists of three core modules, which can be accessed either using its web interface or through shell scripting:
\begin{itemize}\addtolength{\itemsep}{-0.5\baselineskip}

\item {\bf ibootstrap}: This module is reponsible for reading a log file from a lattice computation, extracting the data contained therein, and performing aggregations and statistical analysis. The required inputs to this module are: the name of the log file, the regular expressions required to extract values, the expression to be computed with those values, and optional constraints. Once configured, it loops over these tasks and generates multiple output files in CSV format.
\item {\bf iplot}: This module reads the output of ibootstrap and generates corresponding plots.
\item {\bf ifit}: This module implements various fitting algorithms that can read the output of ibootstrap and cooperate with iplot to include fitting results in plots.
\end{itemize}

The web application is built around these three modules using the web2py framework. The framework provides Application Program Interfaces to perform the following tasks: concurrency handling, generation of dynamic web pages, web services, streaming of large files both in and out, security (authentication, authorization, XSS and Injection prevention), connecting to a database for storage (web2py supports 10 different database engines), maintaining state through user sessions and cookies, caching results of recurrent operations (for speed), and internationalization capabilities.

Plotting is done using the matplotlib library.

\section{Example}

A typical example of medium complexity consists of the computation of a meson's mass. This value can be computed (after Wick rotation to Euclidean time) by fitting a two point correlation
\begin{equation}
C_2(t_x-t_y) \equiv FT_{\mathbf{x}\mathbf{y}} \left< 0 \right|j(\mathbf{x},t_x) j^\dagger(\mathbf{y},t_y)\left|0\right>
\label{f1}
\end{equation}
with a sum of exponentials
\begin{equation}
C_2(t) = \sum_i A_i e^{-m_i t}
\label{f2}
\end{equation}

Here $FT_{\mathbf{x}\mathbf{y}}$ is a zero momentum spatial Fourier transform (in $\mathbf{x}$ and $\mathbf{y}$), and $j(\mathbf{x},t_x)$ is a creation operator with the same quantum numbers as the meson we are interested in. The expectation value is computed numerically on the lattice in Step 2, as descibed in the introduction.

The $m_i$ parameters are the masses of those states with the same quantum numbers as $j$. They can be extracted form the fit. The smaller of mass $m_0$ is the ground state for the meson created by the operator $j$.

Assume we have multiple measures for $C_2(t)$, along with other measurements, one for each gauge configurationm, and that their values are stored in a flat text file as in the following:
\vskip -5mm
\begin{center}
\includegraphics[width=3in]{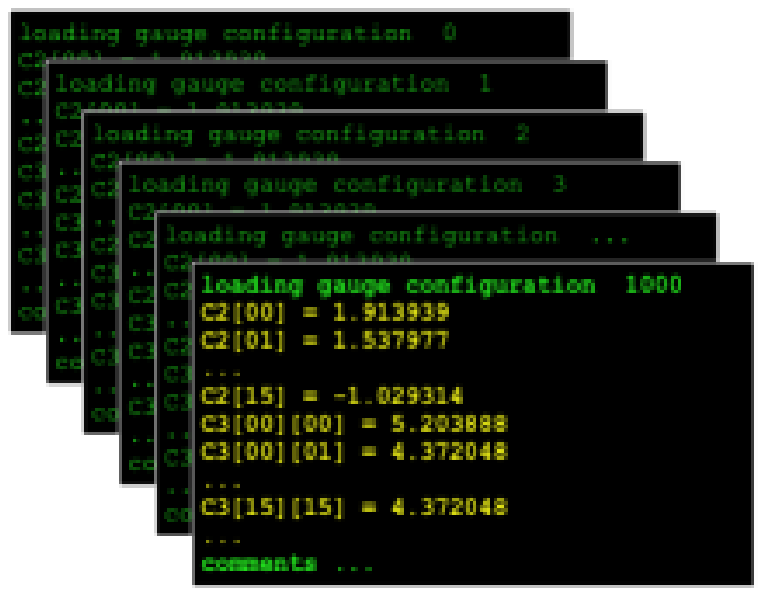}
\end{center}
\vskip -5mm
In this case {\tt C2[i]} is just an arbitrary label -- any name could be used. The index in square brackets is the value for $t$. For each $t$ there are as many measurements as there are gauge configurations. The file may also contain additional text and comments. The different pages in the figure correspond to different gauge congurations and their output is in the same file.

We upload this file into mc4qcd and then ask it to compute the following expression:

\begin{lstlisting}
"C2[<t>]"
\end{lstlisting}

which means: {\it look for all variables labeled \verb!C2[<t>]! where $t$ is an index implicitely defined}. Everything inside double quotes identify variables via their corresponding pattern. Everything outside double quotes must be a valid Python expression, which can use all functions from the Python math module (such as log, exp, sin, etc.), as well as user defined functions.

\begin{figure}
\begin{tabular}{cc}
\includegraphics[width=3in]{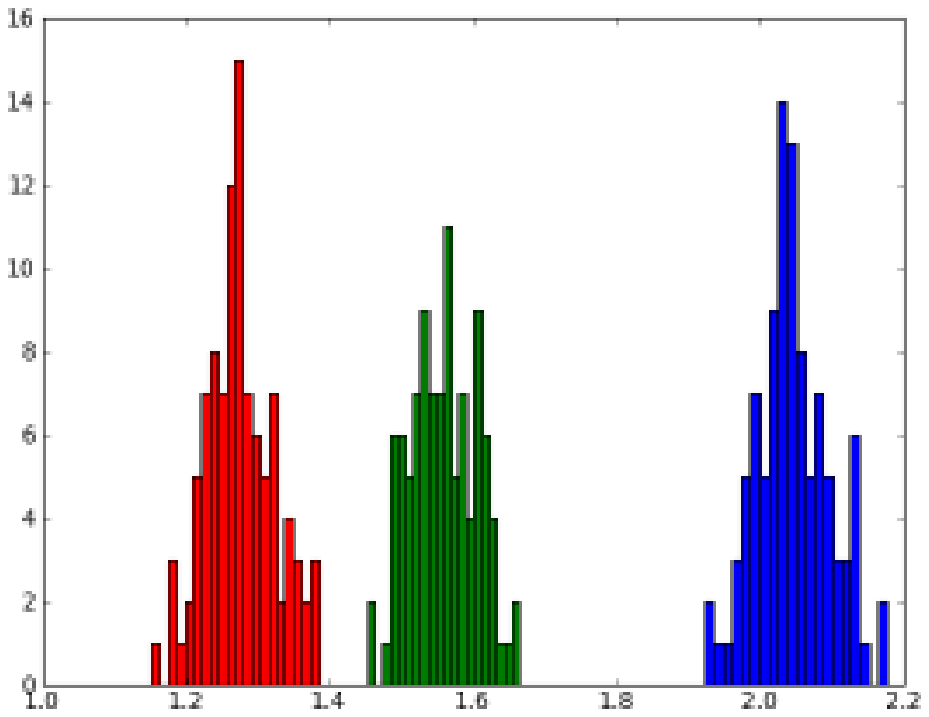} &
\includegraphics[width=3in]{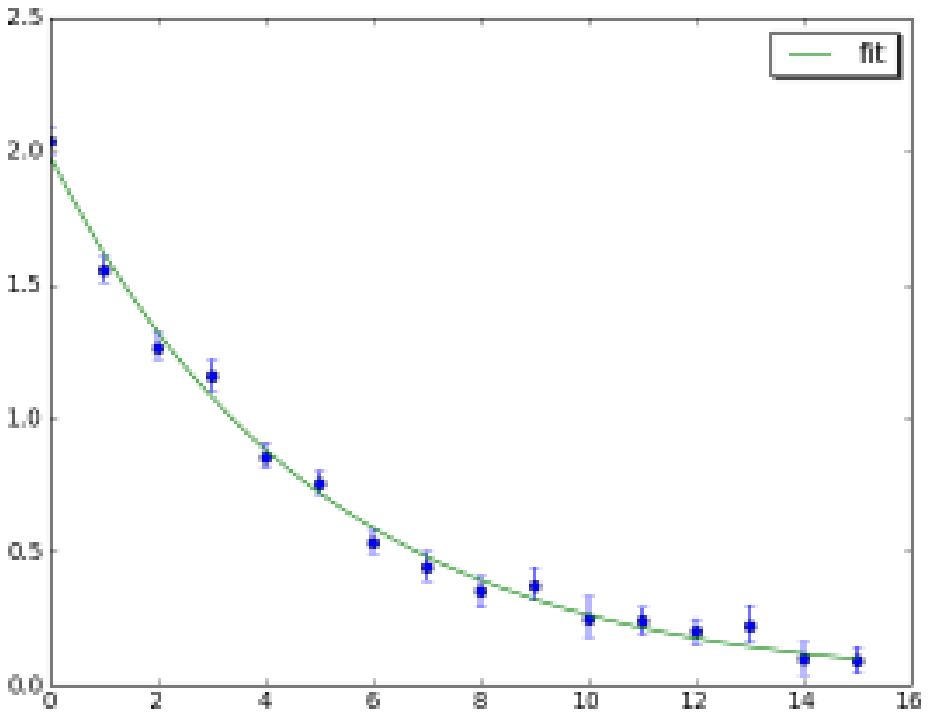}
\end{tabular}
\caption{The figure on the left shows the distribution of bootstrap samples for $C_2(t)$ at three different values of $t$. The figure on the right shows the two point correlation function at different $t$ with its bootstrap errors, fitted by an exponential function.\label{fig2}}
\end{figure}

The program, for each $t$, extracts all values from the file, computes simple averages, moving averages, autocorrelations, bootstrap errors and sample distributions (some sample plots are shown in fig.~\ref{fig2}).

Next, to fit this with an expotential we simply type: 

\begin{lstlisting}
a*exp(-m*t)@a=1,a_bu=0.2,m=0.1
\end{lstlisting}

Here mc4qcd understands that $t$ is the parameter defined implicitly when parsing the data, while $a$ and $m$ are arbitrary unknown parameters that will be determined in the fit. The initial values used for fitting (priors) are specified on the right-hand side of the @ symbol. \verb!a_bu! is the Baysian uncertainly of the prior $a$. In this case, we performed a single mass fit, because this term dominates for large $t$. Multimass fits are also supported by mc4qcd.

The results of the fit are shown, superimposed to the data, in fig.~\ref{fig2}. The fitting results and the Hessian describing the correlation between the fitting parameters ($a$ and $m$) are displayed under the plot.

mc4qcd also allows more complex expression. Here are some examples:

\begin{lstlisting}
log("C2[<t2>]"/"C2[<t2>]") for t1==t2+1
0.5*("C2[<t1>]"+"C2[<t2>]") for t1==32-t2 and t1>16
"C3[<t1>][<t2>]"/"C2[<t1>]"/"C2[<t2>]")
\end{lstlisting}
The latter example extracts a matrix element from a 3 point correlation function $C_3$.

The fitting module also allows for complex expressions such as correlated Bayesian fits:

\begin{lstlisting}
{1:a,2:b}[t1]*exp(-m*t1)@a=1,a_bu=0.2,b=2,m=0.1
\end{lstlisting}
(an exponential fit in $t_1$ with different coefficients depending on $t_2$).

\section{Conclusions}

mc4qcd is a new tool designed spefically for Lattice QCD analysis. It is very general, with certain limitations. For example, even though it can read unstructured data, it still often requires that the data be tagged by strings that can be recognized by the regular expression parser. Planned future enhancements include extending mc4qcd to give it the ability to learn more complex templates and handle more use cases.

Another limitation is the restricted ability to customize plots. Because all of the work is carried out automatically, it generates good quality images for screen presentation, but does not allow as much customization as is sometimes required for publication in scientific journals.

Despite this, mc4qcd can be a valuable tool for increasing collaboration between scientists working in this field and automating some of their daily tasks.

mc4qcd is distributed as a web2py package with a single dependence (the matplotlib library). It can be demoed and donwloaded from \verb!http://latticeqcd.org!

{\bf Acknowledgements}
This work has been supported by the US Department of Energy under grant DEFC02-06ER41441.

\end{document}